\documentclass[reprint,aps, superscriptaddress,nofootinbib]{revtex4-1}
\usepackage{graphicx}
\usepackage{amsmath}
\usepackage{amssymb}
\usepackage{amsfonts}
\usepackage{amsthm}
\usepackage{caption}
\usepackage{color}

\newcommand{\Sch}{{Schr\"{o}dinger }}
\newcommand{\ket}[1]{{|}{#1}{\rangle}}
\newcommand{\bra}[1]{{\langle}{#1}{|}}

\begin{document}

\captionsetup{justification=raggedright,singlelinecheck=false}

\title{Scattering of a composite quasiparticle by an impurity on a lattice}

 \author{Fumika Suzuki}
 \email{fumika@physics.ubc.ca}
 \affiliation{%
Department of Physics, University of British Columbia, Vancouver, V6T 1Z1, Canada
}
\affiliation{%
Institute for Molecular Science, National Institute of Natural Sciences, Okazaki, Japan}

 \author{Marina Litinskaya}
  \email{litinskaya@gmail.com}

 \affiliation{%
Department of Chemistry, University of British Columbia, Vancouver, V6T 1Z1, Canada
}

 \author{William G. Unruh}
   \email{unruh@physics.ubc.ca}
 \affiliation{%
Department of Physics, University of British Columbia, Vancouver, V6T 1Z1, Canada
}

\date{\today}
\begin{abstract}
We study scattering of a composite quasiparticle, which possesses a degree of freedom corresponding to relative separation between two bound excitations, by a delta-like impurity potential on a one-dimensional discrete lattice. Firstly, we show that, due to specific properties of their dispersion, lattice excitations bind to impurities with both negative and positive potentials. We demonstrate that the finite size of the composite excitation leads to formation of multiple excitation-impurity  bound states. The number and the degree of localization of these bound states depend on the signs and relative magnitudes of the impurity potential and the binding strength of two quasiparticles. We also report the existence of  excitation-impurity bound states whose energies are located in the continuum band. Secondly, we study a change in the entanglement between the centre of mass and relative coordinate degrees of freedom of a biexciton wave packet during single impurity scattering and decoherence caused by it.  For a composite quasiparticle on a lattice, the entanglement between its relative and centre of mass coordinate degrees of freedom arises naturally due to inseparability of the two-particle Hamiltonian. One of the main focuses of our study is to investigate how this inseparability affects the creation of the biexciton-impurity bound states and the entanglement dynamics. 
\end{abstract}

\maketitle


\section{Introduction}

Interference of composite objects is an important problem with applications in many areas of physics \cite{arndt, jul, pik}. A composite object possesses internal degrees of freedom, which are often entangled with each other and with external degrees of freedom. This entanglement may act as a source of decoherence in one of the degrees of freedom. Interplay between various degrees of freedom becomes crucial when a composite object -- a wave packet with several entangled internal degrees of freedom -- is split into distinct components by mirrors or other equipment in order to create its spatial superposition state. An important question is how the entanglement among the degrees of freedom changes in the process of wave packet splitting, as it allows one to collect the which-way information on one of the degrees of freedom through measurement of the adjacent degrees of freedom \cite{anton}.

Another important problem is tunnelling of composite objects through barriers \cite{comp2}, with applications in nuclear fusion \cite{comp3}, induced decay of false vacuum \cite{pen'kov}, and tunnelling of Cooper pairs in superconductors and Wannier-Mott excitons in semiconductor heterostructures \cite{shegelski2012}. It had been shown that the probability of tunnelling of an object possessing an internal degree of freedom -- for example, a diatomic molecule -- through a barrier may greatly exceed that of a structureless object with similar properties due to appearance of quasi-bound states in the combined scattering and molecular binding potentials \cite{pen'kov, comp2, comp3}. Furthermore, interaction of a molecule with the external potential can induce transitions between molecular states due to coupling between relative and centre of mass (CM) coordinate degrees of freedom \cite{comp1, shegelski2013}.

Most existing literature study free-space composite objects, e.g., looking at two- or many-photon transport through an impurity \cite{fan, fan2} or impurities \cite{photon},  and extended scattering potentials \cite{pen'kov, comp2, comp3}. Recently, a free-space model with a delta potential has been addressed in \cite{bill}, where the authors considered a diatomic molecule scattered by infinitely narrow mirror. The long-lived scattering resonances and the increase of the entanglement among internal degrees of freedom of the system caused by scattering were reported. While realizing an interaction between a diatomic molecule and a semi-transparent infinitesimally thin mirror is technically demanding in free-space, its lattice analog is available in molecular crystals, where an impurity acts as a delta potential, which scatters collective many-atom excitations -- Frenkel excitons \cite{book}. Scattering of a single exciton by an impurity has been studied in literature \cite{konobeev}. For a two-exciton bound state (biexciton) the problem is more complicated, as recent numerical studies has shown \cite{bulatov}.  Scattering of composite objects in the lattice configuration can be studied also with the help of cold optical systems. Major success in trapping ultracold atoms \cite{cold_atoms} and molecules \cite{cold_molecules} in optical lattices allows for creation of controllable periodic ensembles in many ways similar to natural crystals, which support rotational Frenkel excitons \cite{us, gorshkov}.
The exciton-exciton interactions can be controlled by applying external electric field, and under certain conditions a biexciton is formed \cite{roman}. Perturbing the ideal translational invariance of the lattice ensemble by replacing one of the molecules by a molecule of a different kind simulates an impurity in a natural crystal \cite{us}.

The interaction of two-particle states on a lattice with a defect has been studied previously for the Hubbard \cite{zhang, zhang2, zhang3} and Su-Schrieffer-Heeger \cite{lib} models, with the focus on the possible overlap of a bound state with the continuum and edge bound states.

Here we study scattering of a Frenkel biexciton by an impurity in a one-dimensional (1D) lattice. In our model the on-site interaction for excitations is not a free parameter but corresponds to an infinite repulsion, as one molecule cannot be excited twice. In addition, we have a two-particle interaction between the excitations and we use periodic boundary conditions which impose additional symmetry on the wave function. Some of the continuum models and their results can be obtained from the corresponding lattice model by taking a limiting procedure (e.g. the lattice constant $a\rightarrow 0$, the number of lattice sites $N\rightarrow \infty$) with the excitation hopping strength $J\propto 1/a^2$. However, although many continuum models of two particles interacting via the potential depending only on their relative distance are separable in the relative and CM coordinates, it is often the case that the corresponding lattice Hamiltonian and its eigenstates are no longer separable in those coordinates due to discreteness (Section II). As a result, the entanglement between relative and CM coordinate degrees of freedom can naturally arise for a composite quasiparticle on a lattice, and indeed we will observe that the width of a biexciton wave-function in the relative coordinate depends on the CM wave vector $K$. Moreover, the lattice models have energy bands which are bound both from above and below, that allows the creation of bound states with both attractive and repulsive interactions (Section~\ref{s-effmass}). In the continuum models the energy is bound from below but has no upper bound, and the bound states are only associated with attractive interactions. An important objective of our study is to investigate 
how inseparability of relative and CM coordinates and finiteness of the energy band of a composite quasiparticle on a lattice affects creation of the biexciton-impurity bound states and the entanglement dynamics.
 
The paper is organized as follows: In Section \ref{s-model}, we derive biexciton states analytically in an ideal 1D lattice with periodic boundary conditions. We use them as a basis for the following discussion. In Section~\ref{s-effmass} we study and contrast the eigenstates of an exciton and a biexciton in a 1D lattice with an impurity. We find that the free-space intuition cannot be directly applied to a lattice setup: In particular, binding between the impurity and lattice excitations occurs at both signs of the impurity potential. For an exciton the exact solution is reported. For a biexciton we show numerically that the extra (relative) degree of freedom results in formation of multiple biexciton-impurity bound states, -- in contrast to one exciton, which always has one bound state near a delta-like potential. The number of bound states and the degree of their localization are determined by the signs and relative values of the exciton-exciton and biexciton-impurity interactions. The bound states are also studied analytically by looking at the poles of the scattering amplitude for exciton and biexciton. Furthermore, we report that our model with the impurity can be approximately solved and there exist bound states in the continuum \cite{wigner} in which two excitons are mutually bound and bound to the impurity and the energies of the states are located in the continuum band. In Section \ref{s-wp}, we study scattering of a biexciton wave packet by an impurity, and a change in the entanglement between its relative and CM coordinate degrees of freedom. In Section~\ref{s-discussion} we present our conclusions and discuss further applications of the obtained results.

\section{Biexciton states}
\label{s-model}

We consider a 1D lattice of molecules or any other two-level objects with periodic boundary conditions, with the lattice constant equal to $1$, and study the excitation transfer between the molecules. The Hamiltonian in the nearest-neighbour approximation is 
 \begin{eqnarray}\label{ham}
\hat{H}_0&=&\displaystyle\sum_{n=-N/2+1}^{N/2}(E_0 \hat{a}^{\dagger}_{n}\hat{a}_{n}+ J (\hat{a}^{\dagger}_{n+1}\hat{a}_{n}+\hat{a}^{\dagger}_{n-1}\hat{a}_{n})\nonumber\\
&&\qquad\qquad+D\hat{a}^{\dagger}_{n+1}\hat{a}_{n+1}\hat{a}_{n}^{\dagger}\hat{a}_{n}+L\hat{a}_{n}^{\dagger}\hat{a}_{n}\hat{a}_{n}^{\dagger}\hat{a}_{n})
\end{eqnarray}
where $n$ labels the sites of a 1D lattice, $N$ is the total number of lattice sites, and $n+N$ is taken to be just a different label for the site $n$ for arbitrary $n$. While $N$ could be arbitrary, we take $N$ to be even. The analysis for odd $N$ is possible, but more complicated than for even $N$. Operators $\hat{a}^{\dagger}_{n}$, $\hat{a}_{n}$ describe excitation and de-excitation of $n$-th molecule, $J$ describes the excitation hopping strength between molecules in sites $n$ and $n \pm 1$, while $D$ denotes a two-particle interaction strength between the excitations, and $E_0$ is the one-particle excitation energy. $L\rightarrow \infty$ accounts for the hard-core constraint, i.e., to the fact that one molecule can accommodate at most one excitation. The number operator $\hat{\mathcal{N}}=\sum_{n}\hat{a}^{\dagger}_{n}\hat{a}_{n}$ commutes with the Hamiltonian and the number of the excitations is conserved.

A single {\it exciton} \cite{book} can be represented by the eigenstates of the first two terms of Hamiltonian  (\ref{ham}) when the total number of excitations in the lattice is one. Here we consider Hamiltonian acting on two-exciton subspace, then the basis can be written as $|m, n\rangle=\hat{a}_{m}^{\dagger}\hat{a}_{n}^{\dagger}|0\rangle \in \mathcal{H}_{m}\otimes\mathcal{H}_{n}$, meaning that $m$-th and $n$-th sites are excited. Then $|m,n\rangle\equiv|m,n+N\rangle$, $|m,n\rangle\equiv|m+N,n\rangle$. In addition we have indistinguishability of excitations $|n,m\rangle\equiv|m,n\rangle$ and $|n,n\rangle$ does not exist by the hard-core constraint. We now define $r = n+m$ ($r/2$ is the centre-of-mass coordinate) and relative coordinate of two excitations $s = n-m$ on a lattice. Since the model is integrable for two excitations \cite{choy}, we derive biexciton states analytically in this section. The variables $r$ and $s$ are not independent: they must be both even or odd. This indicates $r+s$ should be even for physical states. However we extend the space of states to all $r$ and $s$ for simplicity, i.e., Hilbert space $ \mathcal{H}_{m}\otimes\mathcal{H}_{n}$ is a subspace of $\mathcal{H}_{r}\otimes \mathcal{H}_{s}$. If we take $-N/2+1 \leq m, n \leq N/2$, then $-N+1\leq r\leq N$ and $-N+1\leq s\leq N$. We define $|r,s\rangle \in  \mathcal{H}_{r}\otimes\mathcal{H}_{s}$ and introduce the unitary transformations $\hat{R}=\sum_{r,s}|r+1,s\rangle\langle r,s|$ and $\hat{S}=\sum_{r,s}|r,s+1\rangle\langle r,s|$. The symmetries of $m,n$ translate to
 $|r,s\rangle\equiv |r+2N,s\rangle\equiv|r,s+2N\rangle \equiv |r+N,s+N\rangle$ for arbitrary $r,s$.
The constraint that $|n,m\rangle\equiv |m,n\rangle$ becomes that $|r,s\rangle\equiv |r,-s\rangle$ so
everything can be taken as defined only for positive $s$. Then, (\ref{ham}) acting on two-exciton subspace is equivalent to
\begin{eqnarray}\label{ham2}
\hat{H}_0&=& 2E_0\displaystyle\sum_{r,s}|r,s\rangle\langle r,s|+J (\hat{R}+\hat{R}^{\dagger}) (\hat{S}+\hat{S}^{\dagger})\nonumber\\
&&+D\displaystyle\sum_{r,s}\delta (|s|-1)|r,s\rangle\langle r,s|+L\displaystyle\sum_{r,s} \delta (s)|r,s\rangle\langle r,s|\nonumber\\
\end{eqnarray}

Note that the Hamiltonian only couples even $r+s$ states with each other, and odd $r+s$ states with each other. So one can always project the solutions back onto the $|m,n\rangle$ set of states. 

We consider the wave-function $\Phi (r,s)$ such that the state is
\begin{eqnarray}
|\Phi\rangle =\displaystyle\sum_{r,s}\Phi(r,s)|r,s\rangle
\end{eqnarray}
and $\Phi(r,s=0)=0$ arises from the limit of finite energy when $L\rightarrow \infty$.

The periodic boundary conditions and indistinguishability of excitations impose the following symmetry requirements:\footnote{We have $(2N)^2$ states in $ \mathcal{H}_{r}\otimes\mathcal{H}_{s}$. $\Phi(r+N,s\pm N)\equiv\Phi(r,s)$ divides it by $2$, eliminating $s=0$ subtracts $2N$, $\Phi(r,s)\equiv\Phi(r,-s)$ divides it by $2$ and even $r+s$ divides it by $2$. This process gives $N(N-1)/2$ states, which is the number of states in $ \mathcal{H}_{m}\otimes\mathcal{H}_{n}$.}
\begin{equation}\label{symmetries}
\begin{array}{cc}
\Phi (r,s) & \equiv \Phi (r+2N,s) \equiv \Phi (r,s+2N) \\

\\

& \equiv \Phi (r+N,s\pm N) \equiv \Phi (r,-s).
\end{array}
\end{equation}

The Hamiltonian commutes with $\hat{R}$ and we can simultaneously diagonalize $\hat{R}$ and $\hat{H}_0$. The eigenvalues of $\hat{R}$ must be pure phases. Since $\hat{R}^{2N}=\hat{R}$, they can be written as $e^{iK}$ with the eigenstates
\begin{equation}
\Phi (r,s) = e^{i K r} \phi_K (s), \hskip 1cm K = \frac{2\pi l_K}{2N},
\end{equation}
where $l_K \in [-N+1, N]$ is an integer. The last two symmetry requirements in (\ref{symmetries}) indicate
\begin{eqnarray}
\phi_{K} (s)=(-1)^{l_{K}}\phi_{K} (N-|s|)
\end{eqnarray}
which means that the relative coordinate wave-function $\phi_K (s)$ is even or odd about $s = \pm N/2$ (as we assume $N$ is even, $N/2$ is an integer) according to the parity of $l_K$:
\begin{equation}\label{even or odd}
\phi_K (N/2+|s|) = (-1)^{l_K} \phi_K (N/2-|s|).
\end{equation}

Projecting the eigenvalue equation $\hat{H}_0|\Phi \rangle = E|\Phi \rangle$ onto $\langle r,s|$ away from $s=0, \pm 1$, we get
\begin{eqnarray}\label{eigenvalueeq0}
2J\cos K (\phi_{K} (s-1)+\phi_{K} (s+1))=(E-2E_0) \phi_{K} (s)\nonumber\\
\end{eqnarray}
which yields
\begin{eqnarray}\label{symantisym}
\phi_{K}(s) = \begin{cases}
 \cos k (N/2 - |s|), \mbox{ if } l_{K} \mbox{ is even},\\
 \sin k (N/2 - |s|),  \, \mbox{ if }  l_{K} \mbox{ is odd}\\
\end{cases}
\end{eqnarray}
up to normalization. The corresponding energy eigenvalue is given by 
\begin{eqnarray}\label{twoexcitonenergy}
E=2E_0+4J\cos K \cos k.
\end{eqnarray}

The eigenvalue equation at $s=1$ gives
\begin{eqnarray}\label{eigenvalueeq}
2J \cos K \phi_{K} (2)+D\phi_{K}(1)=4J\cos K\cos k\phi_{K} (1).
\end{eqnarray}

Then substituting (\ref{symantisym}) into (\ref{eigenvalueeq}), we have
\begin{eqnarray}\label{inteq}
\frac{D}{2J\cos K}&=&\frac{\cos kN/2}{\cos k (N/2-1)}, \mbox{ if } l_{K} \mbox{ is even},\nonumber\\
\frac{D}{2J\cos K}&=&\frac{\sin kN/2}{\sin k (N/2-1)}, \,\mbox{ if } l_{K} \mbox{ is odd}.
\end{eqnarray}

In this paper, we are interested in the bound two-exciton complex, {\it biexciton} \cite{vektaris, spano, roman}, which appears as a result of the exciton-exciton interactions given by the $D$ and by
the $L$ terms in the Hamiltonian. For biexciton, the wave-function of the relative coordinate decays exponentially with the growth of separation and $k$ is complex, i.e., $k = k_r + i k_i$. As the two-particle energies are proportional to $\cos k$, we conclude that $k_r = 0$ or $\pi$ to keep the biexciton energy real. Now a transformation $K\rightarrow K\pm\pi$ and $k\rightarrow k \pm\pi$ gives
\begin{eqnarray}
&&e^{iKr}\cos k (N/2-|s|)\rightarrow e^{i (K\pm \pi)r}\cos (k\pm \pi)(N/2-|s|)\nonumber\\
&&=(-1)^{r}e^{iKr}\cos (k(N/2-|s|)\pm \pi (N/2-|s|))\nonumber\\
&&=(-1)^{r}e^{iKr}(\cos k (N/2-|s|)\cos  \pi (N/2-|s|)\nonumber\\
&&\quad \qquad\qquad\quad \mp\sin k (N/2-|s|)\sin  \pi(N/2-|s|))\nonumber\\
&&=(-1)^{r+s}e^{iKr}\cos k (N/2-|s|)
\end{eqnarray}
when $l_{K}$ and $N$ are even. When $l_{K}$ is odd and $N$ is even, similarly we obtain $e^{iKr}\sin k (N/2-|s|)\rightarrow (-1)^{r+s} e^{iKr}\sin k (N/2-|s|)$. 

Furthermore, it can be confirmed that if $K$, $k$ obey the equation (\ref{inteq}), then $K\pm\pi$, $k\pm\pi$ also do. Note that these arguments are valid for both real and complex $k$, i.e., they are valid for the biexciton states as well as for continuum two-exciton states.  For even $N$, the wave-function for $K\pm\pi$, $k\pm\pi$ is the same as that for $K$, $k$ up to the factor $(-1)^{r+s}$ which is equal to $1$ on the even $r+s$ sublattice. When $N$ is odd, these transformations become more complicated, and it is for this reason that we chose $N$ to be even throughout this paper. Eq.(\ref{inteq}) with complex $k$ shows that the condition for biexciton to appear is $|D| \geq |2J \cos K|$. When $|D/2J|>1$, one has a biexciton solution for each value of $K$. If we choose $|D/2J|>1$ and large $N$, then the equation can be simplified to
\begin{equation}\label{conditions}
\frac{D}{2J\cos K} \approx e^{-i(k_r + i|k_i|)}, \hskip 1cm k_r = 0 \ {\rm or}\ \pi.
\end{equation}

Here the choice of $k_r$ correlates with the sign of $D/2J \cos K $. In particular, if sgn($J$) = sgn($D$), the solution of (\ref{conditions}) exists if: (1a) $k_r = 0$, $-\pi/2 \leq K \leq \pi/2$, or (1b)  $k_r = \pi$, $\pi/2 \leq |K| \leq \pi$. Similarly, if sgn$(J) \neq$ sgn($D$), the solution exists if: (2a) $k_r = 0$, $\pi/2 \leq |K| \leq \pi$, or (2b)  $k_r = \pi$, $-\pi/2 \leq K \leq \pi/2$. However, as discussed above, we have $\Phi_{K\pm\pi}(r,s) =  (-1)^{r+s} \Phi_{K}(r,s)$ for even $N$ and linearly dependent states can be summed as $\frac{1}{2}(|\Phi_{K}\rangle+|\Phi_{K\pm\pi}\rangle)=\frac{1}{2}(1+(-1)^{r+s})|\Phi_{K}\rangle$, which is zero for odd $r+s$ states and $|\Phi_{K}\rangle$ for even $r+s$ states. In the following sections we use these states setting sgn($J$) = sgn($D$), $k_r = 0$, and $K \in (-\pi/2, \pi/2]$ \footnote{For sgn$(J) \neq$ sgn($D$), either the $K$-domain or the value of $k_r$ should be modified.}.

Eq.(\ref{conditions}) gives 
\begin{eqnarray}\label{k_i}
k_i \approx \left| \ln \left| \alpha_K  \right| \right| \quad \mbox{with} \quad \alpha_{K}=\frac{2J\cos K}{D}.
\end{eqnarray}

Substituting $k_i$ into (\ref{twoexcitonenergy}), we can write the biexciton energy as
\begin{equation}\label{E_b}
E_{b} (K)\approx 2E_0+D(1+\alpha^2_{K}).
\end{equation}

 The biexciton wave-function can now be written as \footnote{For even $r+s$ sublattice, the sum can be taken over $\{ r,s \}$ having the same parity, and $s\neq 0$ and positive: $\sum_{\{r,s\}} = \sum_{s = -N+1}^{N-1} \sum_{r = -N+2+|s|}^{N-|s|}$ with step 2 in the sum over $r$.}
\begin{eqnarray}\label{Phi_K}
|\Phi_{K}\rangle =\sqrt{\frac{2}{N}}\displaystyle\sum_{r,s}e^{iKr}\phi_{K} (s) |r,s\rangle
\end{eqnarray}
and
\begin{equation}\label{phi_K}
\begin{array}{cc}
\phi_{K} (s) =\frac{1}{{\cal N}_e }\cosh k_i (N/2-|s|) , & l_K \ {\rm even},\\

&\\

\phi_{K}(s) =\frac{1}{{\cal N}_o}  \sinh k_i (N/2-|s|) , & l_K \ {\rm odd},\\
\end{array}
\end{equation}
where the normalization constants are ${\cal N}_{e,o} = \sqrt{N-1 \pm \sinh k_i(N-1)/\sinh k_i}$ (upper sign for ${\cal N}_e$, lower for ${\cal N}_o$). Since $k_{i}$ is related to $K$ by (\ref{k_i}), the biexciton wave-function can be expressed just in terms of $K$.

We note that $k_{i}$ goes to infinity as $K \to \pi/2$. Since (\ref{phi_K}) is maximum at $|s|=1$, $N-1$, all values except those are infinitely smaller. Then after normalization, biexciton wave-function is represented by delta functions at $K=\pi/2$:
\begin{equation}
\displaystyle
\phi_{K=\pi/2}(s) = \frac{1}{2}(\delta (|s|-1) + (-1)^{l_{K}}\delta (N-|s|-1)).
\end{equation}

Note that $\phi_{K}(s)$ is defined on $0<s<N$ and extended to other values by applying the symmetries (\ref{symmetries}). In the above expressions and the corresponding normalizations, we already extended $\phi_{K}(s)$ to $-N<s<N$ with the symmetry $\phi_{K}(s)=\phi_{K} (-s)$. Also we have $\phi_{K}(s=0)=0$. 

We use the biexciton wave-function derived here as a basis in the following sections.

\section{Interaction of excitons and biexcitons with impurity}\label{s-effmass}

We now assume that an impurity is located at the origin of the lattice. In this section we study the biexciton-impurity bound state(s), and Section~\ref{s-wp} discusses scattering of a biexciton wave packet by an impurity potential. Note that, in both sections, we assume a biexciton is tightly bound in its relative coordinate with $|D/2J|\gg1$, and transitions from biexciton states to continuum two-exciton states and vice versa are not considered.

\begin{figure}
{%
\includegraphics[clip,width=1\columnwidth]{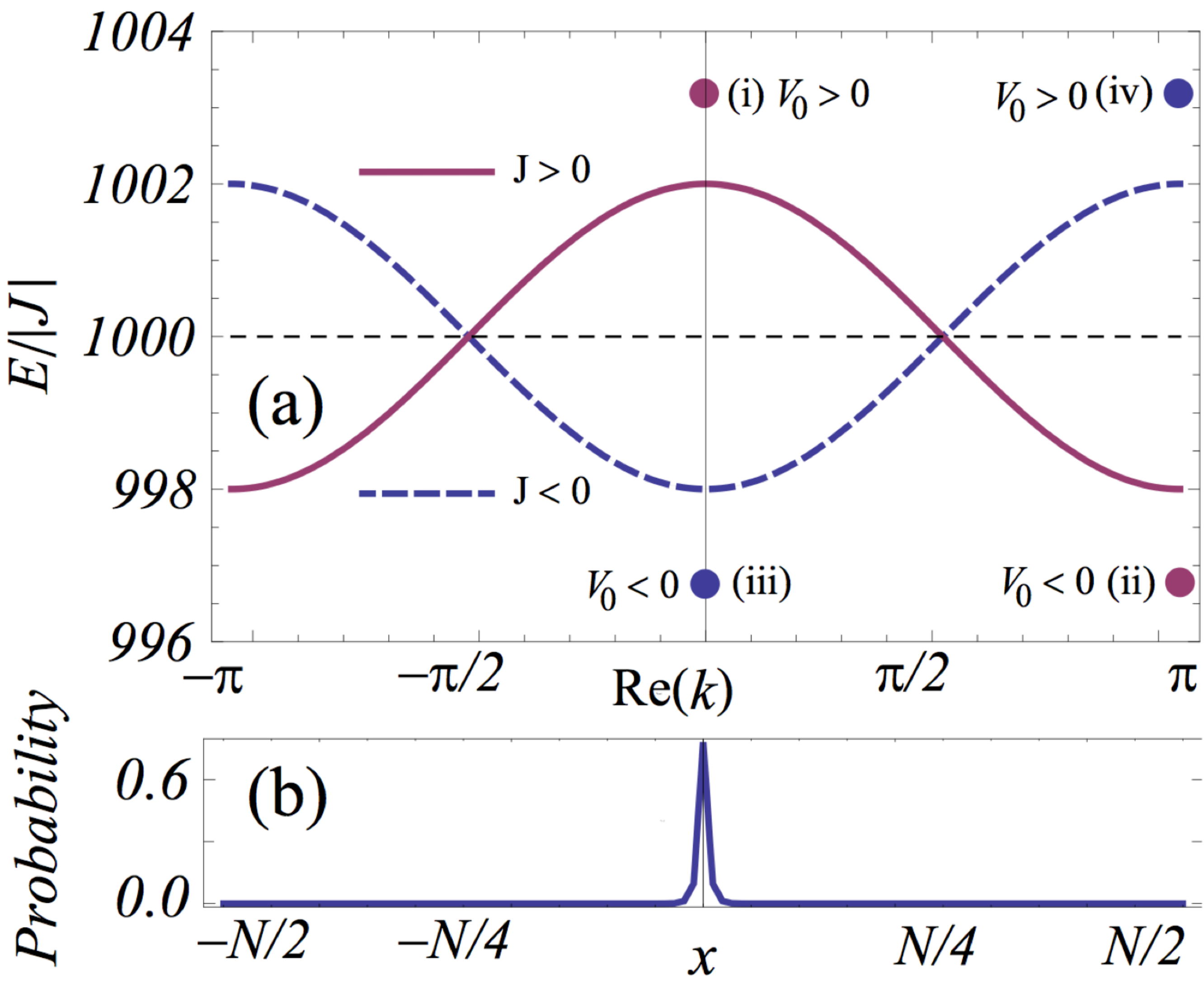}%
}
\caption{(Colour online) Single exciton interacting with impurity: Single bound state. (a) Energy spectrum for various combinations of $J$ and $V_0$. $J>0$: Continuum states (red solid line) and bound states (i) or (ii); $J<0$: Continuum states (blue dashed line), and bound states (iii) or (iv). (b) Wave-function corresponding to bound states (i-iv).}\label{f-exc}
\end{figure}

\subsection{Exciton-impurity interaction: Single bound state}\label{exactexciton}

As a benchmark, let us consider interaction of a single exciton with an impurity. This problem was addressed in \cite{konobeev} with a parabolic approximation for the exciton dispersion. Here we show that the account of the convex-concave dispersion of the exciton leads to qualitatively new behaviour of exciton-impurity binding.

In the nearest-neighbour approximation, the Hamiltonian is 
\begin{eqnarray}\label{excitonNNA}
\hat{H} &=&  \displaystyle J \sum_{n} \biggl( \ket{n}\bra{n-1} + \ket{n-1}\bra{n} \biggr)+V_0\displaystyle\sum_{n}\delta (n)|n\rangle\langle n|\nonumber\\
&&+E_0\displaystyle\sum_{n}|n\rangle\langle n|
\end{eqnarray}
where $|n\rangle=\hat{a}^{\dagger}_{n}|0\rangle$ and $V_0$ is the impurity strength, which is equal to the difference in the excitation energies of the impurity and the host molecules.

For the periodic boundary conditions, exciton states without an impurity can be written as
\begin{equation}\label{P(k)}
|\varphi (k)\rangle =\frac{1}{\sqrt{N}}\sum_{n}\varphi_{k}(n)|n\rangle ,\quad \varphi_{k}(n)=e^{ikn}
\end{equation}
where $k = 2 \pi \nu / N$ is the wave vector, and $\nu \in [-N/2+1, N/2]$ is an integer. 

The eigenstates for an exciton interacting with an impurity in a 1D lattice can be found exactly. Using (\ref{P(k)}), we divide the eigenstates into antisymmetric and symmetric ones:
\begin{equation}
\begin{array}{c}
|\varphi_a(k_a)\rangle=\sqrt{\frac{2}{N}}\displaystyle\sum_{n}\sin k_an \ket{n},\\

\\

|\varphi_s(k_s)\rangle = \frac{1}{\sqrt{N}}\displaystyle\sum_{n}(\cos k_sn +\alpha \sin k_s|n| )\ket{n},
\end{array}
\end{equation}
where $\alpha$ is yet unknown constant to be determined from the boundary conditions. 

The projection of the eigenvalue equation $\hat{H} |\varphi_{a/s}\rangle = E |\varphi_{a/s}\rangle$ onto a state $\langle n|$ gives for arbitrary $n\not= 0$
\begin{equation}\label{E_e}
E_e(k) = E_0 + 2J \cos k.
\end{equation}

The antisymmetric states $|\phi_a(k_a)\rangle$ vanish at the impurity location $n=0$, so they are  impurity-free states; $k_a = 2\pi \nu_{a}/N$, $\nu_{a}$ is an integer in the interval $[-N/2+1, N/2]$. The states $|\phi_s(k_s)\rangle$, in contrast, interact with the impurity. The projection of the eigenvalue equation onto the state $\langle n=N/2|$ with the account of the periodicity requirement $N/2+1 \to -N/2+1$ gives the equation, which connects $\alpha$ and $k_s$:
\begin{equation}\label{alpha0}
\alpha = \tan \frac{k_s N}{2},
\end{equation}
and the projection of the eigenvalue equation onto the state $\langle n=0|$ gives the remaining equation relating $\alpha$ and $k_s$ with $V_0$:
\begin{equation}\label{alpha}
\alpha \sin k_s = - V_0/2J.
\end{equation}

From (\ref{alpha0}) and (\ref{alpha}) we conclude that the wave vectors $k_s$ obey the following equation:
\begin{equation}\label{k_s}
\tan \frac{k_s N}{2} \sin k_s = - \frac{V_0}{2J}.
\end{equation}

Its solutions tend to $k_s = 2\pi n_s/N$ when $V_0 \to 0$, and the parameter $V_0/2J$ determines their shifts from the impurity-free values. All solutions but one are real values corresponding to exciton-impurity scattering states. The energies of the scattering states $E_e(k_{s,a})$ are plotted in FIG 1 (a) as function of the wave vector for $J > 0$ (red solid line) and $J < 0$ (blue dashed line); $E_0/|J|=1000$, $|V_0|/|J|=2.5$.

One solution of (\ref{k_s}), $k=k_b \equiv k' + i k''$ with ${\rm Re}(k)=k'$, ${\rm Im}(k)=k''$ is complex, and describes exciton bound to the impurity. Interestingly, the structure of $k_b$ is determined by the relative signs of $V_0$ and $J$. As the energy is real, $\cos k_b = \cos k' \cosh k'' - i \sin k' \sinh k''$ must be real as well, which means that $k'$ equals to either 0, or $\pi$. As follows from (\ref{k_s}), the first case is realized when $V_0/2J > 0$, the second -- when and $V_0/2J < 0$. For large $N$ we can write the wave vector for the bound state as the simplified form:
\begin{equation}\label{ex_poles}
\begin{array}{cccc}
{\rm sgn}(V_0) = {\rm sgn}(J): & & k' = 0, & \displaystyle k'' = {\rm arsinh}\frac{V_0}{2J};\\

&&&\\

{\rm sgn}(V_0) \neq {\rm sgn}(J): & & k' = \displaystyle \pi, & \displaystyle k'' = - {\rm arsinh}\frac{V_0}{2J}.\\
\end{array}
\end{equation}

Therefore, the spectrum always possesses a single bound state with the energy $E_e(k_b)$. For a given sign of $J$, it forms both for negative and positive impurity potentials. In the first case the bound state splits downwards, in the second -- upwards from the continuum band. If the signs of $J$ and $V_0$ coincide, the state is described by purely imaginary wave vector (cases (i) and (iii) in FIG.~\ref{f-exc}~(a)). If the signs of $J$ and $V_0$ are different, its wave vector has a real part $\pi$ (cases (ii) and (iv)). The wave-function of the bound sates looks the same for all four situations (i-iv) (FIG. 1 (b)), as if a bound state forms under repulsive forces exactly as it does for attractive forces. In fact, bound complexes ``forming under repulsive forces" in a lattice geometry have recently been in the focus of attention \cite{win, he}. As we show below, this equivalence between attractive and repulsive potentials allows for a simple interpretation in terms of the effective mass $m_{\rm eff}^{-1} = (\partial^2 E_e(k)/\partial k^2)/\hbar^2$ of the exciton.

Indeed, the effective masses defined in the centre ($\mbox{Re}(k) \sim 0$) and at the edge ($\mbox{Re}(k) \sim \pi$) of the exciton energy band have different signs, owing to the convex-concave dispersion of the exciton. Due to the structure of the \Sch equation, $\hat{H} \to -\hbar^2 \Delta / (2 m_{\rm eff}) + \hat{V}(n)$, a particle with negative effective mass ``sees" an attractive potential as repulsive, and repulsive as attractive \cite{eco}. For concreteness of the following discussion, assume $J>0$. Then $m_{\rm eff}^{\rm centre} < 0$, and $m_{\rm eff}^{\rm edge} > 0$ (for $J < 0$ it will be the other way around). Then $V_0 > 0$ will be felt as an attractive potential by the states with $k \approx 0$, while $V_0 < 0$ will be felt as an attractive potential by the states with $k \approx  \pi$, see FIG~1~(a). This interpretation is confirmed by the location of the states at the $(k,E)$-plane obtained analytically, see (\ref{ex_poles}). Similar logic works for $J < 0$, with the flipping of the sign of $V_0$. We conclude that states (ii) and (iii) correspond to attraction of a positive-mass particle by a negative potential, the states (i) and (iv) -- to attraction of a negative-mass particle by a positive potential.

\subsection{Biexciton-impurity interaction: Multiple bound states}\label{bimpsec}

Here we turn to biexciton scattering by an impurity, and find numerically the eigenstates of time-independent Schr\"{o}dinger equation $\hat{H}|\Psi\rangle = E|\Psi\rangle$, where $\hat{H}=\hat{H}_0+\hat{V}$ with $\hat{V}=V_0\sum_{r,s}(\delta (r+s)+\delta (r-s))|r,s\rangle\langle r,s|$\footnote{$\hat{V}=V_0\sum_{m,n}(\delta(n)+\delta (m))|m,n\rangle\langle m,n|=V_0\sum_{r,s}(\delta(r+s)+\delta(r-s))|r,s\rangle\langle r,s|$ shows that only even $r+s$ states interact with an impurity.}. Note that the impurity potential $\hat{V}$ also obeys periodic boundary conditions of $|r,s\rangle$. We expand $|\Psi\rangle$ in the basis of the free-biexciton wave-functions as $|\Psi^{\mu}\rangle =\displaystyle\sum_{K} u^{\mu} (K)|\Phi_{K}\rangle$, where $\mu$ is the state index and the biexciton wave-function $|\Phi_{K}\rangle$ is given in (\ref{Phi_K}). By choosing just the biexciton states as a basis, our results are approximate,
which should be good as long as $|D/2J|\gg1$ (i.e., the energy of the biexciton states
is much larger than any of the unbound states). We have:
\begin{equation}\label{matrix}
\displaystyle\sum_{K'}\langle \Phi_{K}|\hat{V}|\Phi_{K'} \rangle u^{\mu}(K') = (E^{\mu} - E_b(K)) u^{\mu} (K),
\end{equation}
where $\langle\Phi_{K}|\hat{V}|\Phi_{K'}\rangle=V_{KK'}$ is written as
\begin{eqnarray}\label{Veff}
V_{KK'}= \frac{4V_0}{N}
\displaystyle\sum_{\substack{s\not= 0\\ (r+s \, {\rm even})}}\phi^{*}_{K}(s)\phi_{K'}(s) e^{i(K'-K)s}.
\end{eqnarray}

\begin{figure}
{%
\includegraphics[clip,width=1\columnwidth]{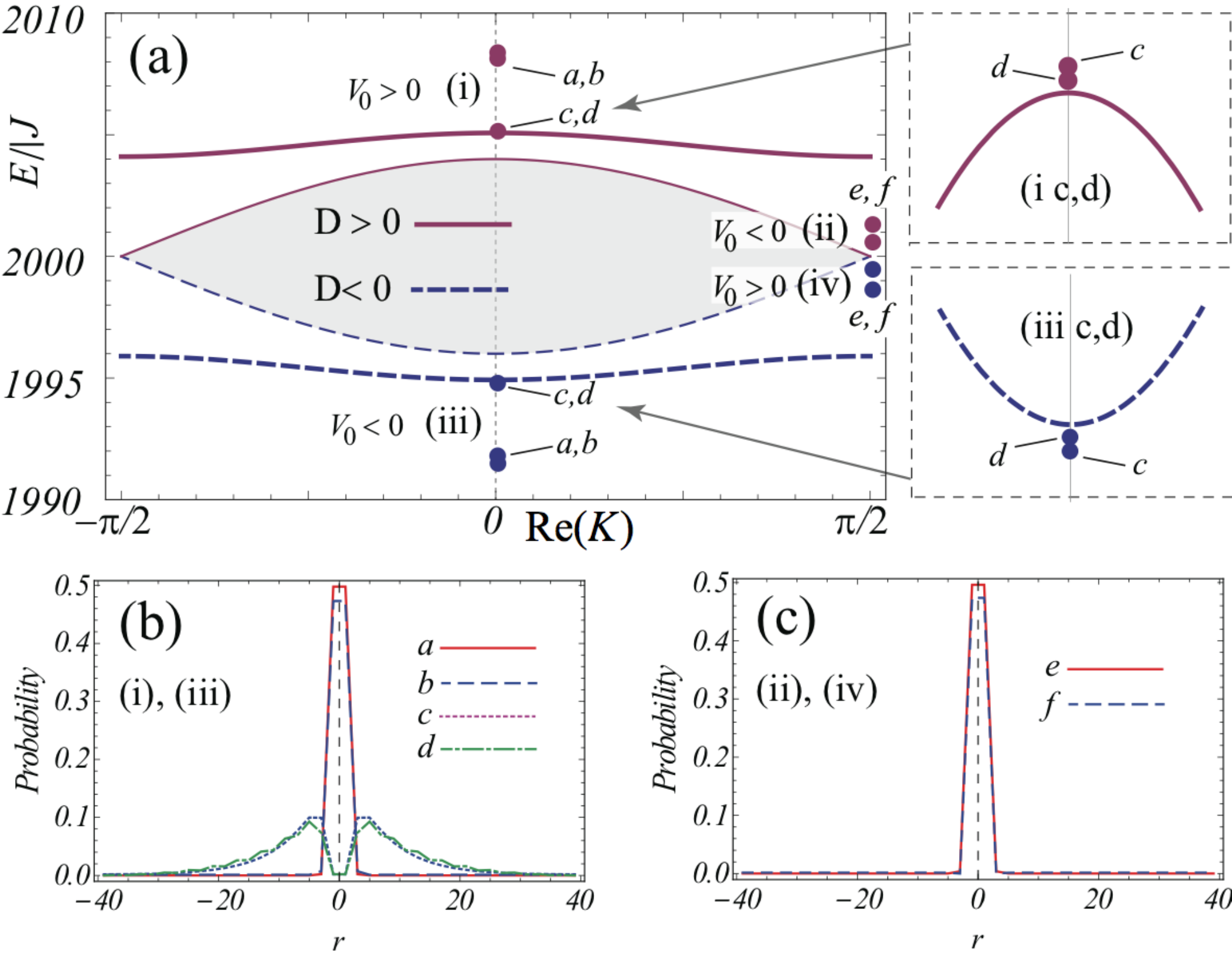}
}
\caption{ (Colour online). Biexciton interacting with an impurity: Multiple bound states. (a) Biexciton scattering states for $D>0$ (red solid line) and $D<0$ (blue dashed line), bound biexciton-impurity states (dots (i,ii) for $D>0$ and (iii,iv) for $D<0$), and two-exciton unbound states (grey shaded region). Right panels zoom the regions (i) and (iii) with multiple states. (b) The probability distributions of the bound states (i,iii), and (c) of the bound states (ii,iv).}\label{f-biexc}
\end{figure}

The eigenvalues of biexciton-impurity bound states can be obtained by numerical diagonalization of the matrix $M_{KK'}=E_{b}(K)\delta_{KK'}+V_{KK'}$ and they are shown in FIG 2 (a) for $N=40$, $E_0/|J|=1000$, $|V_0|/|J|=4$, $|D|/|J|=4.1$ as blue and red dots. All combinations of signs of $D$ and $V_0$ are considered. As follows from (\ref{E_b}), the effective mass of biexciton is defined by the sign of $D$, and the sign of $J$ is irrelevant, so we set sgn($J$) = sgn($D$). For reference, thick solid lines show the impurity-free biexciton dispersion, the grey shade shows two-exciton unbound states, and the isolated points -- the biexciton-impurity bound states. Their positions on the Re($K$)-axis (near Re($K$) $ = 0$ for cases (i) and (iii) and near Re($K$) $ =  \pi/2$ for cases (ii) and (iv)) is justified by the analogy with the exciton case, and will be proved analytically below, see ~(\ref{level_bi}). Note that although $K$ is not a good quantum number in the interaction region, we can use it as a quantum number in the asymptotic region, i.e., it is a good quantum number in the region away from the scattering centre.

Near Re($K$) $\sim 0$ we see four -- marked as $a, b, c, d$ -- isolated eigenvalues of the type (i) for $D>0, V_0>0$ (attraction of a negative-mass quasiparticle by a positive potential), and of type (iii) for $D<0, V_0<0$ (attraction of a positive-mass quasiparticle by a negative potential). The bound character of these states is confirmed by the decaying shape of their probability distribution in $r$-coordinate at $s=1$ (FIG~2~(b)); At larger $s$ the wave-functions have same behaviour. States $a$ and $b$ are well-split from the continuum and are strongly localized near the impurity. The states $c$ and $d$ lie very close to the continuum and are loosely bound to the impurity. In turn, near Re($K$) $\sim \pi/2$ we see two isolated eigenvalues -- they are marked by $e$ and $f$ -- of the type (ii) for $D>0, V_0<0$ (attraction of a positive-mass quasiparticle by a negative potential), and of type (iv) for $D<0, V_0>0$ (attraction of a negative-mass quasiparticle by a positive potential). Their probability distribution in $r$-coordinate at $s=1$ is shown in (FIG~2~(c)). We conclude that the interplay between the biexciton binding and impurity potential leads to formation of multiple bound states with various degree of excitation localization.

The appearance of additional bound states and the variation in their number for different combinations of ${\rm sgn}(D/V_0)$ can be explained by averaging of the scattering potential by the relative coordinate of two bound excitons. We draw the analogy with the work \cite{narrowing}, which examines the averaging of the interface roughness potential in semiconductor heterostructures by electron-hole relative coordinate in a Wannier-Mott exciton. As a result of a composite structure of this exciton, the correlation length of the effective potential acting on the exciton centre of mass greatly exceeds a typical scale of the initial disorder potential, being of the order of the electron-hole mean separation. In a similar way, ~(\ref{Veff}) shows that the relative coordinate of two excitations, which is described by $\phi_K(s)$, acts onto the CM coordinate of the biexciton as an effective potential $V^{\rm eff}_{KK'} (s) = V_0 \phi^{*}_{K}(s)\phi_{K'}(s)$. This can be seen by analogy with the problem of one structureless particle in an extended potential $\hat{V} = \sum_m V_{*}(m) \ket{m} \bra{m}$, whose matrix element in the wave vector space is $V_{k k'} = \frac{1}{N} \sum_m V_*(m) e^{i(k-k')m}$. The spatial extent of the effective potential is determined by the spread of involved $\phi_{K}(s)$ and $\phi_{K'}(s)$. We checked that for small values of $D$ the wave-function $\phi_{K\sim\pi/2}(s)$ is much narrower than $\phi_{K\sim 0}(s)$. Then near Re($K$) $\sim 0$, the averaging is effective and the potential is similar to a well of finite width, while near Re($K$) $\sim\pi/2$ the effective potential is close to the underlying delta function. For large $D$ the wave-function $\phi_{K\sim0}(s)$ narrows due to stronger exciton-exciton binding. Accordingly, for parameters $(D,V_0)$ and $(D,-V_0)$ we expect strong asymmetry in the number of bound states for small $D$, and the same number of bound states when $D$ is large.

\begin{figure}
{%
\includegraphics[clip,width=1\columnwidth]{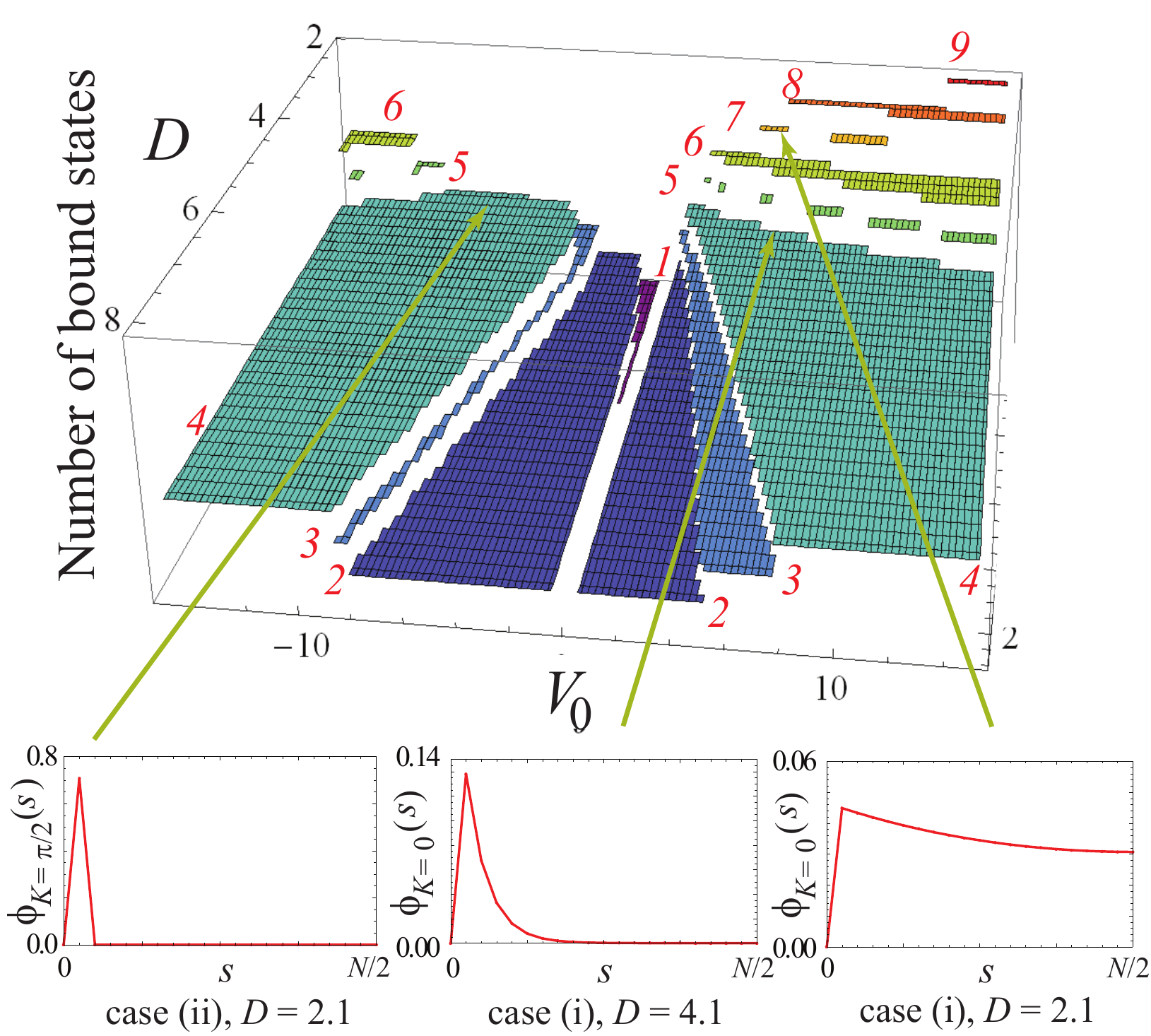}%
}
\caption{(Colour online) Number of biexciton-impurity bound states as function of parameters. All energies are in units of $|J|$. The number of states depends on the width of $\phi_{K}(s)$. The bottom plots show the wave-functions of relative coordinate for three values of parameters; ``case (i)" and ``case (ii)" refer to the notations of FIG.~\ref{f-biexc}~(a). First and last bottom plots illustrate the difference in the width of $\phi_K(s)$ at $K = 0$ and $K = \pi/2$ for same values of all other parameters.}\label{f-number-of-states}
\end{figure}

This is confirmed by examining the number of bound states as function of $(D/J, V_0/J)$ summarized in FIG.\ref{f-number-of-states}. The obtained numbers of bound states are marked as red italic numbers near each plateau in $(V_0,D)$-plane. The state was considered as bound if its energy fell out of the impurity-free biexciton energy band, and its amplitude was a decaying function of $r$. The averaging of the potential by the relative coordinate is illustrated by showing the profiles of three wave-functions: $\phi_{K = \pi/2}(s)$ for $D = 2.1J$, $\phi_{K = 0}(s)$ for $D = 4.1J$, $\phi_{K = 0}(s)$ for $D = 2.1J$. Each plot is associated with the corresponding point in $(D,V_0)$-plane by an arrow. The left graph is for $V_0 = -5|J|$, the middle and the right -- for $V_0 = 5|J|$. Indeed, large numbers of bound states are achieved only with $|D| \leq 2.1 |J|$ and sgn($D$) = sgn($V_0$), when the biexciton is weakly bound and the bound state forms near Re($K$) $ = 0$. However, at small values of $D$ biexciton is resonant with the continuum two-exciton states, and the scattering of biexciton into two-exciton continuum and into the states with one free exciton, and one exciton bound to the impurity may become important.

\subsection{Poles of the reflection amplitude}

The bound states can be examined analytically as the poles of the scattering amplitude with complex momentum which can be derived from the Lippmann-Schwinger equation \cite{scattering}. In this section, we compare the poles of the scattering amplitude derived from Lippmann-Schwinger equation with the assumption $N\to \infty$, and numerical results for (\ref{matrix}). Treating $k = k' + i k''$, we find that the scattering amplitude of an exciton with the dispersion $E_e(k) = E_0 + 2J \cos k$ scattered by an impurity can be calculated (Appendix A) as
\begin{eqnarray}\label{Re}
\displaystyle
\mathcal{R}_e(k) &=& V_0~\biggl[ \biggl( 2J\cos k'\sinh |k''| - V_0 \biggr)  \nonumber\\
 &&\displaystyle- 2iJ\sin k'\cosh |k''| \biggr]^{-1}.
\end{eqnarray}

Note that the exact solution of the problem without the assumption of $N\to \infty$ is already given in Section \ref{exactexciton} and the poles of (\ref{Re}) appear at specific values of $k$ given by (\ref{ex_poles}).

For biexciton the Lippmann-Schwinger equation $|\Psi\rangle=|\Phi_{K}\rangle+\hat{G}_0\hat{V}|\Psi\rangle$ is more complicated, as the potential for the interaction between the impurity and biexciton is non-separable. We solve it approximately using method of continued fractions \cite{sasakawa} in the first-order (see Appendix B). The scattering amplitude with complex $K = K' + i K''$ is found as (see Appendix B):

\begin{equation}\label{Rb}
\begin{array}{c}
\mathcal{R}_b(K) =  2D V_0 S(K',|K''|) \biggl[ \biggl( J^2 \cos (2K') \sinh (2|K''|) - \\

\\

2D V_0 S(K',|K''|)\biggr) -  i J^2 \sin (2K') \cosh (2|K''|) \biggr]^{-1},\\
\end{array}
\end{equation}
where
\begin{equation}
S(K) = \sum\limits_{s = -N/2+1}^{N/2} e^{-2|K''|s} \phi_{K'-i|K''|}(s) \phi_{K'+i|K''|}(s).
\end{equation}

The factor $V_0 S(K)$ accounts for averaging of the potential by the wave-function of relative coordinate, $\phi_K(s)$. The scattering amplitude (\ref{Rb}) has poles for $K' = 0$ and $\pi/2$, with the corresponding equation for $K''$ being
\begin{equation}\label{level_bi}
\begin{array}{c}
\displaystyle \sinh (2|K''_{\rm pole}|) = \frac{2D V_0 S(K'_{\rm pole}, |K''|_{\rm pole})}{J^2 \cos(2K'_{\rm pole})},\\

\\

\displaystyle K'_{\rm pole} = 0, \frac{\pi}{2}.
\end{array}
\end{equation}

This shows that the small parameter of the perturbative expansion is $2DV_0/J^2$. The first-order approximation allows us to confirm the above picture of potential averaged by the wave-function of relative coordinate. It also reproduces the same effect as for a single exciton: when $D$ and $V_0$ have same signs, the bound state appears near $K'=0$, while when $D$ and $V_0$ have opposite signs, the bound state appears near $K'=\pi/2$. The physical meaning of this effect, as can be seen from FIG~2~(a), again lays in the different signs of the effective mass of the biexciton at $\mbox{Re}(K)\sim 0$ and at $\mbox{Re}(K)\sim\pi/2$. We remark that the numerical studies of Ref.~\cite{bulatov} report anomalously high transmission at $\mbox{Re}(K) = \pi/2$ (in our notations, which corresponds to $\mbox{Re}(K) = \pi$ in the notations of Ref.~\cite{bulatov}), and attribute it to existence of a two-body resonant localized state. The authors considered negative $J$ and $D$ and positive $V_0$. Our results confirm this interpretation. Comparison of ~(\ref{level_bi}) and (\ref{ex_poles}) reveals that the bound states at Re($K$) $=\pi/2$ are a general property of a restricted (cosine) energy band, rather than being due to the composite character of the scattered biexciton.

\begin{figure}
\includegraphics[clip,width=1\columnwidth]{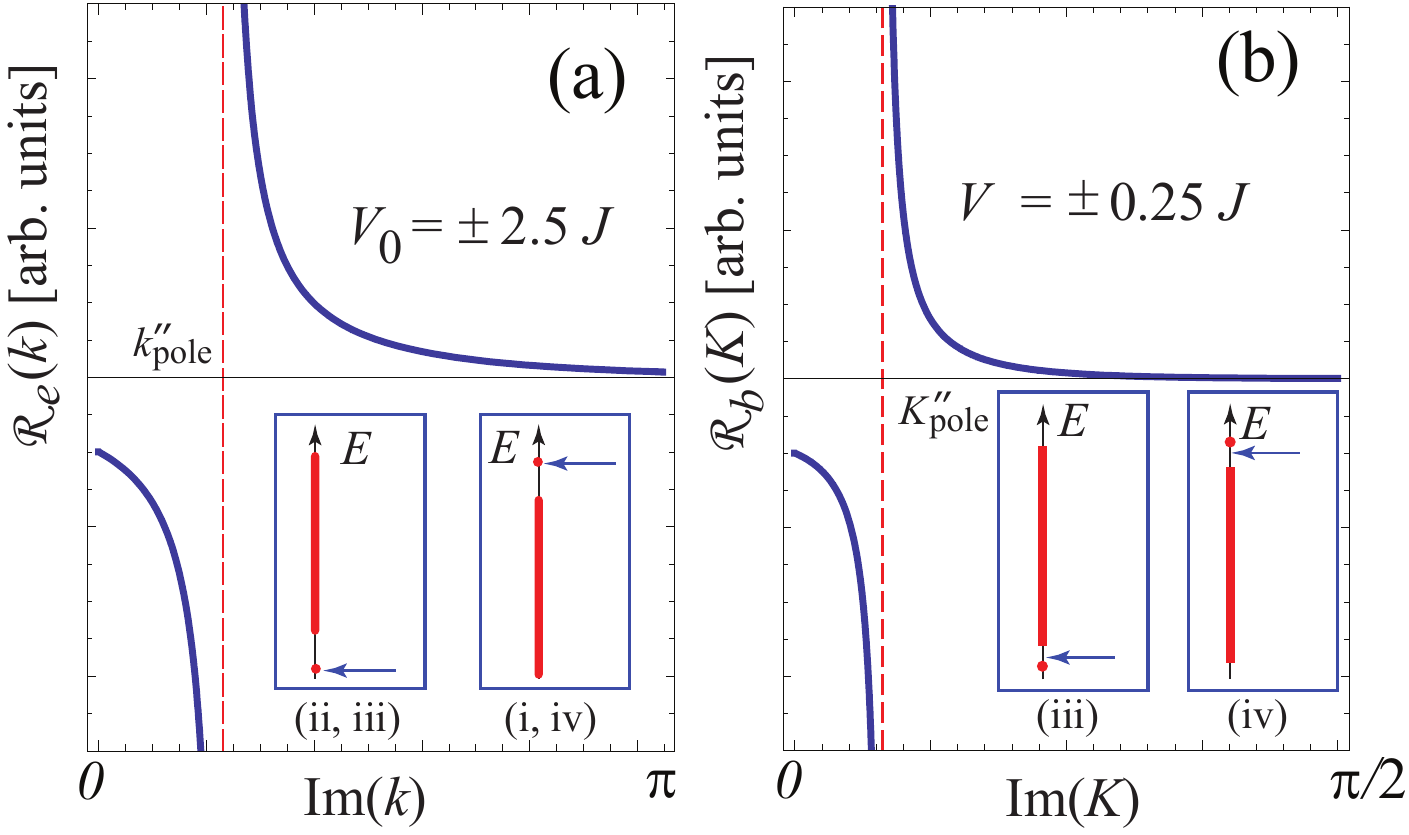}
\caption{(Colour online). Scattering amplitude of a single exciton (a) and biexciton in the perturbative limit (b) as function of complex momentum. The insets show the continuum spectrum (red stripe) and the single bound state (red dot) calculated numerically, and the arrow indicates the analytical estimates (see text).}\label{f-scattering amplitudes}
\end{figure}

In FIG~\ref{f-scattering amplitudes} we show the scattering amplitudes: for an exciton as a function of $k''$ (\ref{Re}), and for a biexciton as a function of $K''$ (\ref{Rb}) with the following parameters: $V_0 = \pm 2.5 J$ for panel (a), $D = 4 J$ and $V_0 = \pm 0.25 J$ (so that we remain in the perturbative limit) for panel (b). Both scattering amplitudes show a single pole, $\mathcal{R}_e$ at $k''$ given by (\ref{ex_poles}), and $\mathcal{R}_b$ at $K''$ given by (\ref{level_bi}). The insets show numerically calculated energy bands, with the bound states shown as dots above or below the continuum. The arrows indicate, respectively, the position of $E_e(k'_{\rm pole} + ik''_{\rm pole})$ and $E_b(K'_{\rm pole} + iK''_{\rm pole})$. For exciton the agreement is perfect, for biexciton - approximate.

\subsection{Bound states in the continuum}

In the previous section, we focused on biexciton-impurity bound states whose energies are located outside the continuum band. In this section, we show that our model can also have bound states in the continuum in which two excitons are mutually bound and bound to the impurity and the energies of the states are located in the continuum band. Although an impurity destroys integrability of the model \cite{and}, it was observed that the Bose-Hubbard model with a particle-particle interaction and an impurity potential in the two-particle sector is semi-integrable in some cases \cite{zhang, zhang2, zhang3}. Our model has $L\rightarrow \infty$ that accounts for the hard-core constraint, therefore $m$, $n$ translations do not commute with the Hamiltonian even if a two-particle interaction $D$ and an impurity potential $V_0$ are zero. However, the model can be approximately solved as follows.

We assume that the strength of a two-particle interaction $D$ and that of an impurity potential $V_0$ are not similar, i.e., either $|V_0| \ll |D|$, or $|V_0| \gg |D|$ (the condition $|D| > 2|J|$ for biexciton formation is always assumed). Under each of these conditions, potentials in the CM and relative coordinate decouple, and the two-particle wave function looks as if $D$ and $V_0$ independently create bound states in the relative and the CM coordinate, respectively. When they are similar, i.e., $|V_0| \sim |D|$, often each exciton individually forms a bound state with a combined potential of $D$ and $V_0$ in $m$ and $n$ coordinates respectively rather than in the relative and the CM coordinates. For simplicity, we do not consider this situation and derive only an approximate solution under the assumption that the strong exciton-exciton interaction ($D$) creates a bound state in the relative coordinate, and weak exciton-impurity interaction ($V_0 \ll D$) perturbatively bounds biexciton to the impurity. Alternatively, each of two excitons strongly bounds to the impurity, and weak exciton-exciton interaction ($D \ll V_0$) forms a biexciton.

Then an ansatz for a two-exciton state with a two-particle interaction $D$ and an impurity potential $V_0$ can be approximately written as 
\begin{eqnarray}\label{ansatz}
|\Phi_{K}^{a}\rangle &\approx&\frac{2}{\sqrt{N}}\displaystyle\sum_{r,s}\sin (K_{a} r) \phi_{K_{a}} (s)|r,s\rangle.
\end{eqnarray}

Here we write the ansatz as a product of wave-functions in the CM and in the relative coordinate since we assume that these two degrees of freedom decouple because of incomparable scales for $V_0$ and $D$. The wave vector $K$ bears index $a$ as we will consider states which are antisymmetric under the replacement $r \to -r$. The problem involving the states which are symmetric under
this transformation becomes more complicated as in the case of  \cite{zhang, zhang2, zhang3}. Given that $\phi_{K}(s)$ is symmetric under $s\rightarrow -s$, $|\Phi_{K}^{a}\rangle $ is antisymmetric under the transformation $(r,s)\rightarrow (-r,-s)$. 

Now we consider the eigenvalue equations. We have four cases: (i) for $|r|\not=|s|$ and $s\not=0, \pm1$ where neither $V_0$ nor $D$ appears, (ii) for $|r|\not=|s|$ and $s=\pm1$ where $D$ appears but $V_0$ does not, (iii) for $|r|=|s|$ and $s\not=0, \pm1$ where $V_0$ appears but $D$ does not, and (iv) for $|r|=|s|$ and $s=\pm1$ where both $D$ and $V_0$ appear. The eigenvalue equations corresponding to (i) and (ii) are already given in (\ref{eigenvalueeq0}) and (\ref{eigenvalueeq}) respectively.

For (iv), we project the eigenvalue equation $\hat{H}|\Phi_{K}^{a}\rangle \approx E|\Phi_{K}^{a}\rangle$ onto $\langle r,s|$ at $r=s=1$:
 \begin{eqnarray}\label{eigen4}
 &&2J\cos K_{a}\phi_{K}(2)+(D+V_0)\phi_{K_{a}}(1)\nonumber\\
 &&\approx4J\cos K_{a}\cos k \phi_{K_{a}}(1).
 \end{eqnarray}
 
 Ideally we look for the state that satisfies all equations corresponding to cases (i), (ii), (iii) and (iv). However, (\ref{eigen4}) reduces to  (\ref{eigenvalueeq}) when $|D|\gg |V_0|$, and it reduces to the eigenvalue equation corresponding to (iii) which is shown below when $|V_0| \gg |D|$. Therefore, in either of these two limits, we only derive the state that satisfies three equations corresponding to (i), (ii) and (iii).  Eqs. (\ref{eigenvalueeq0}) and (\ref{eigenvalueeq}) are already solved in Section \ref{s-model}, and it was found that $k$ satisfies (\ref{k_i}) when the state is bound in the relative coordinate due to a two-particle interaction $D$. Then only thing left is to evaluate complex $K_{a}$ that gives the bound state in the CM coordinate due to the impurity potential $V_0$.

For (iii), we project the eigenvalue equation $\hat{H}|\Phi_{K}^{a}\rangle \approx E|\Phi_{K}^{a}\rangle$ onto $\langle r,s|$ at $r=s$ (here we choose $r=s=N/2-1$), which leads to
\begin{eqnarray}\label{antisymeq}
V_0 \approx 2J\left(\cos K_{a}-\frac{\sin K_{a}}{\tan K_{a} (N/2-1)}\right)\cos k
\end{eqnarray}
where $K_{a}$ is complex for the bound state in the CM coordinate.

When we have $K_{a}=iK''_{a}$ or $K_{a}=\pi/2+iK''_{a}$ and large $N$, (\ref{antisymeq}) reduces to
\begin{eqnarray}\label{antisymeq2}
 V_0/2J &\approx& \pm e^{-K''_{a}}\cosh k_{i}, \quad K_{a}=iK''_{a},\nonumber\\
 -V_0/2J &\approx& \pm e^{K''_{a}}\cosh k_{i}, \quad K_{a}=\frac{\pi}{2}+iK''_{a},
\end{eqnarray}
where we have $+$ sign and $-$ sign on the right hand side when $k=ik_{i}$ and $k=\pi +ik_{i}$ respectively. Here we substituted complex $k$ given by (\ref{k_i}) obtained from  (\ref{eigenvalueeq}) so that the state is bound in both CM coordinate with complex $K$ and in relative coordinate with complex $k$ respectively. When $K_{a}=iK''_{a}$, the equation with $+$ sign corresponds to the case where $\mbox{sgn}(J)=\mbox{sgn} (D)=\mbox{sgn} (V_0)$, and that with $-$ sign corresponds to the case where $\mbox{sgn}(J)\not= \mbox{sgn} (D)=\mbox{sgn} (V_0)$. On the other hand, when $K_{a}=\pi/2+iK''_{a}$, the equation with $+$ sign corresponds to the case where $\mbox{sgn} (J)=\mbox{sgn} (D)\not=\mbox{sgn} (V_0)$, while that with $-$ sign corresponds to the case where $\mbox{sgn} (J)=\mbox{sgn} (V_0)\not=\mbox{sgn} (D)$. Substituting $K_{a}$ obtained by (\ref{antisymeq2}) into (\ref{twoexcitonenergy}) gives
\begin{eqnarray}\label{impenergy}
E_{b1}&=&\frac{DV_0 (D+ V_0- \sqrt{4J^2+(D- V_0)^2})}{2(DV_0- J^2)},\nonumber\\
E_{b2}&=&\frac{DV_0 (D + V_0+ \sqrt{4J^2+(D- V_0)^2})}{2(DV_0- J^2)}\nonumber\\
\end{eqnarray}
where both equations in (\ref{antisymeq2}) give the same result and $2E_0$ should be added if $E_0$ is not set to equal to zero.

We find that $E_{b1}$ can fall into the continuum band that covers the interval $[2E_0-4J,\mbox{ }2E_0+4J]$ while $E_{b2}$ falls outside of the band when $D, V_0>0$. On the other hand, $E_{b2}$ can fall into the continuum band and $E_{b1}$ falls outside of it when $D, V_0 <0$. Therefore, one of them corresponds to the type of bound states which we discussed in the previous section, while the other is a bound state in the continuum similar to that obtained in  \cite{zhang, zhang2}.

\begin{figure}
{%
\includegraphics[clip,width=1\columnwidth]{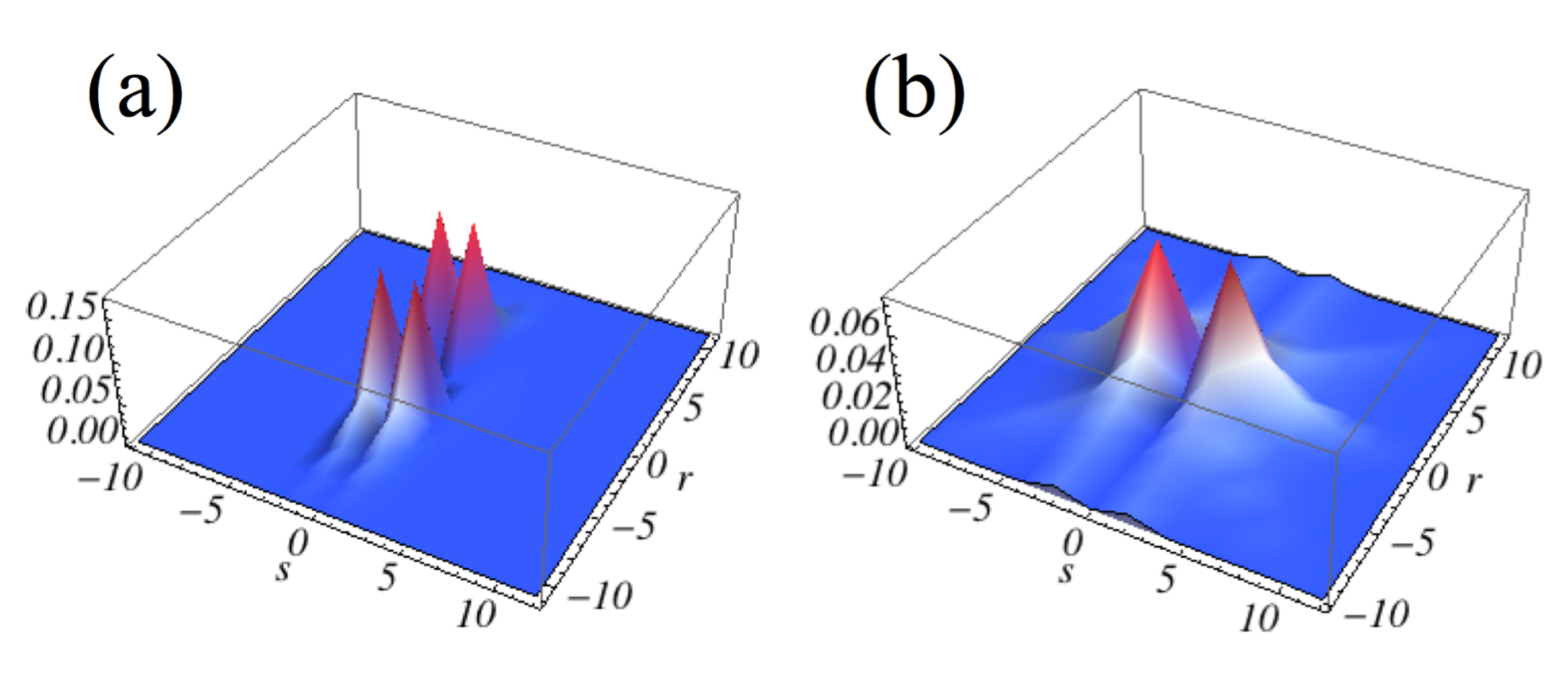}%
}
\caption{(Colour online). Bound states in the continuum. (a) $D=4.1 J$, $V_0=8J$. (b) $D=4.1J$ , $V_0=J$.}\label{biexcitonimp}
\end{figure}

We have also studied numerical solutions of the Schr\"{o}dinger equations in the full basis, which included all continuum states as well biexciton states discussed in previous sections. We find four distinct types of bound states: the first type is the biexciton state where two particles are bound in the relative coordinate with complex $k$ but the biexciton complex is not bound by the impurity (free biexciton). The second type corresponds to two particles whose CM position is bound by the impurity with complex $K$ but they are not bound in the relative position (two excitons whose CM position is bound to the impurity). The third type of bound states corresponds to the case where one particle is bound by the impurity (the wave vector of one particle is complex) but the other is not. Finally, the forth type of bound states corresponds to two particles mutually bound and bound to the impurity. In this forth case, the energies of the bound states are approximately given by Eq. (\ref{impenergy}), and one of them falls into the continuum. FIG. \ref{biexcitonimp} shows an example of a bound state in the continuum of the forth type. In the limit $D \ll V_0$, we found almost exact agreement between $E_{b1}$ in (\ref{impenergy}) and the eigenvalue of the state obtained by numerical diagonalization. Indeed we see the probability distribution of the state demonstrates clear decoupling between $r$ and $s$ (FIG. \ref{biexcitonimp} (a)), which indicates that the ansatz (\ref{ansatz}) written as a product of wave-functions in the CM and in the relative coordinate works well. Meanwhile, in the limit $D\gg V_0$, we found that the formula for $E_{b1}$ works only approximately, and the probability distribution shows that full decoupling between $r$ and $s$ is not achieved (FIG. \ref{biexcitonimp} (b)).

\section{Decoherence by internal degrees of freedom}
\label{s-wp}

\begin{figure*}
 \includegraphics[clip,width=2\columnwidth]{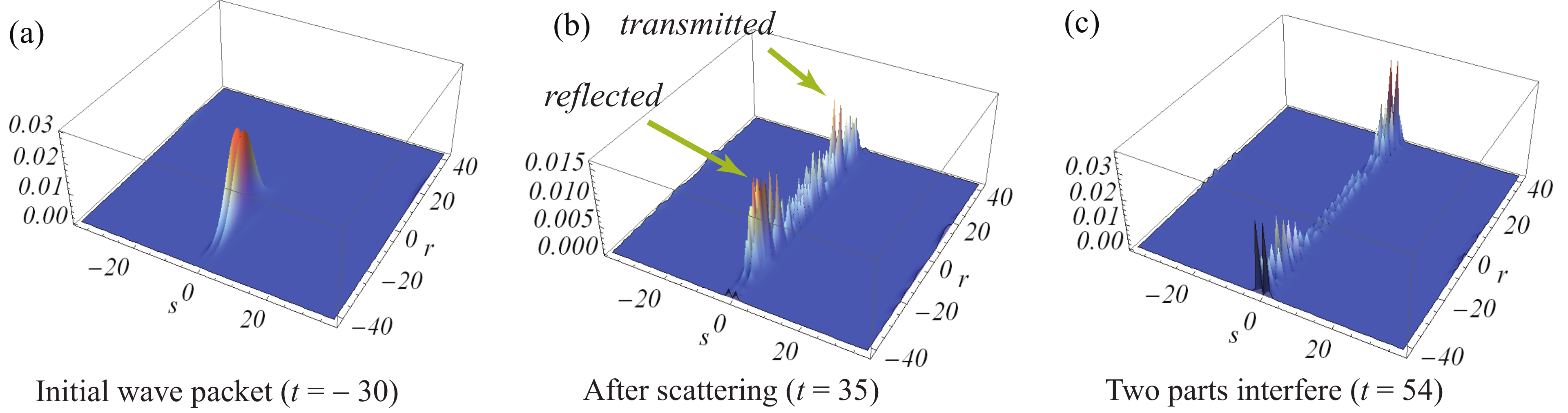}
\caption{(Colour online). Time evolution of the biexciton wave packet (\ref{wp-initial}) interacting with the impurity. Shown is the probability distribution $\rho (r,s; r,s; t)$, with $N = 40$, $r \in (-N, N)$, $s \in (-N, N)$.}\label{f-wp-dynamics}
\end{figure*}

In this section we study the change in the entanglement between the CM motion and the internal degree of freedom $s$ via scattering by an impurity. We consider a single scattering of a narrow biexciton wave packet $\sum_K u_K | \Phi_K \rangle$ with the expansion coefficients $u_{K}$ by an impurity. Let $W(r, s) = \sum_K u_K f_K(r) \phi_K(s)$ where $f_{K}(r)=e^{iKr}$ and $\phi_{K} (s)$ is given in (\ref{phi_K}) denote its real-space projection onto a state $| r, s \rangle$. The reduced density matrix of $\rho(r, s; r', s') = W(r, s) W^*(r', s') $ is
\begin{equation}\label{rhocm}
\begin{array}{ll}
\rho_{cm} (r,r') & = \sum\limits_{s} \rho(r, s; r', s)\\
&\\
&= \sum_{K, K'}u_{K'}^* u_K f_{K'}^*(r') f_{K}(r) \sum\limits_{s} \phi_{K'}^*(s) \phi_K(s).\\
\end{array}
\end{equation}

Therefore, tracing over the relative coordinate degrees of freedom suppresses the contribution of the pairs of components with different $|K|$; The less similar $\phi_K$ and $\phi_{K'}$ functions are, the stronger the suppression. In a sense, the relative coordinate degrees of freedom acts as a source of decoherence. To quantify this decoherence, we consider a thought experiment, in which the two ends of a lattice are connected (ring geometry). A wave packet propagates towards the impurity and is split by it into transmitted and reflected parts. The two parts propagate away from the impurity, meet at the opposite side of the ring and interfere. The off-diagonal elements of $\rho_{cm}$ describing this interference of the wave packet with itself quantify the degree of decoherence.

We solve the time-dependent Schr\"{o}dinger equation, $i\hbar \frac{\partial}{\partial t}|\Psi (t)\rangle =\hat{H}|\Psi (t)\rangle$, where $\hat{H}=\hat{H}_0+\hat{V}$ as in the previous section. In the basis set of static impurity-free biexciton wave-functions (\ref{Phi_K}), $|\Psi (t)\rangle = \sum_{K} u_{K}(t) |\Phi_{K} \rangle$, the time-dependent expansion coefficients $u_{K} (t)$ of the wave vector $K$ obey equations of motions given by
\begin{equation}\label{uKeq}
i\hbar \frac{\partial u_{K}(t)}{\partial t}=\displaystyle\sum_{K'}\langle\Phi_{K}|\hat{H}|\Phi_{K'}\rangle u_{K'} (t)
\end{equation}
where $\langle\Phi_{K}|\hat{H}|\Phi_{K'}\rangle =M_{KK'}=E_{b}(K)\delta_{KK'}+V_{KK'}$ is given in (\ref{Veff}).

Here we project the Hamiltonian onto the biexciton set of states, but ignore two-exciton continuum states. This simplification is only possible when the biexciton state is well split from the two-exciton continuum with $|D|>4|J|$ and the impurity potential $V_0$ is not large compared to $D$, i.e., not $|V_0| \gg |D|$. However, when $|D|\sim 2|J|$, a scattering by an impurity can destroy biexciton states and make them decay into two-exciton continuum states. In this paper, we do not study such physical process, but show only one simple example in which a biexciton wavepacket scatters by an impurity without the transitions between biexciton and two-exciton continuum states.

We consider an initial wave packet of the form
\begin{equation}\label{wp-initial}
| \Psi_0 \rangle = \frac{1}{\mathcal{N}} \displaystyle\sum_{K} u_{K}(0) | \Phi_K \rangle,
\end{equation}
where $\mathcal{N}$ is the normalization factor, and
\begin{equation}\label{uK(0)}
u_{K}(0)=e^{-\frac12 (K-K_0)^2 / (\Delta K_0)^2}.
\end{equation}

We solve a matrix differential equation (\ref{uKeq}) numerically with an initial condition (\ref{uK(0)}).  We choose the parameters as $K_0 = 3\pi/8$, $\Delta K_0 = \pi/24$, $D/J=4.5$, $|V_0/J| \sim |D/J|$ such that half of the wave packet is transmitted and half of it is reflected in the CM coordinate. Here $D/J$ and $V_0/J$ have opposite signs. The calculations are done with the number of molecules $N=40$, and $2E_0$ chosen as a reference point for energy. We measure time in units of $1/|J|$. The wave packet shown in FIG.~\ref{f-wp-dynamics}~(a) starts at $t=-30$ and is split by the impurity at around $t=0$ into reflected and transmitted parts (FIG.~\ref{f-wp-dynamics}~(b)). Finally, two parts meet at the opposite side of the ring and interfere with each other ($t=54$, FIG.~\ref{f-wp-dynamics}~(c), shows the moment of maximal overlap).

The von Neumann entropy \cite{newmann}, that measures the entanglement between CM and relative coordinate degrees of freedom, is defined as
$
S(t) = -\operatorname{tr} [\rho_{cm}(r, r'; t)\log_{2}\rho_{cm}(r, r'; t)],
$ 
and
$\rho_{cm}(r, r'; t) = \sum_{s} \Psi (r,s; t)\Psi^{*} (r',s; t)$ where $\Psi (r,s;t)=\sum_{K}u_{K}(t)f_{K}(r)\phi_{K} (s)$. We diagonalize $\rho_{cm} (r,r';t)$ as $\rho_{cm}=\sum_{r}\eta_{r}|r\rangle\langle r|$ and evaluate $S=-\sum_{r}\eta_{r}\log_{2}\eta_{r}$. The entropy for the initial biexciton wave packet [(\ref{wp-initial}) and FIG.~\ref{f-wp-dynamics}~(a)] is calculated as $S = 0.18$. In contrast to the case of the free-space composite object \cite{bill}, the lattice Hamiltonian (1) is not separable into relative and CM coordinates. As a result, the $r-s$ entanglement of a state changes when it propagates. Careful choice of the wave packet parameters can make this change negligibly small for long enough propagation times. For our parameters $\Delta S \sim 10^{-2}$ from $t=-30$ to around $t=0$. However, when the wave packet is scattered by the impurity at $t \sim 0$, the interplay between the exciton-exciton interaction and the impurity potential changes the entropy rapidly to $S = 0.38$. Note that it is likely that a larger change in the entropy can be observed when the transitions between biexciton and
two-exciton continuum states occur.

\begin{figure*}
\includegraphics[clip,width=2\columnwidth]{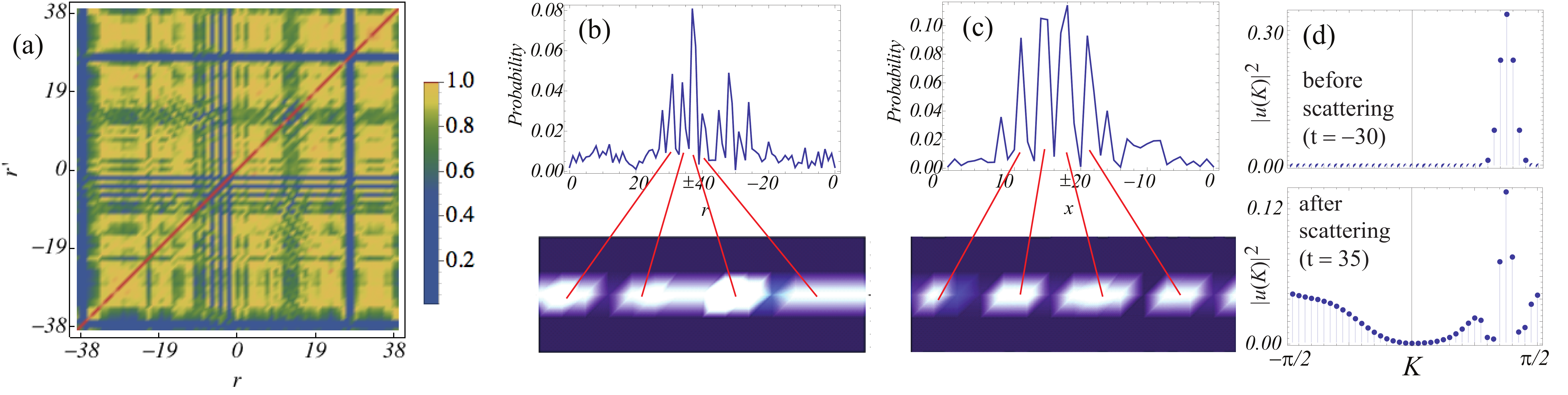}
\caption{(Colour online). (a) Comparison of diagonal and off-diagonal matrix elements for the reduced density matrix after scattering, $C(r,r')$ at $t = 35$. (b) Interference pattern produced by the reduced density matrix in the CM-coordinate at the moment of maximal overlap, $\rho_{cm}(r, r; t=54)$. (c) Similar calculation for single exciton (see text). (d) Mode distribution before and after scattering. Expectation of energy $\langle E_{b}\rangle=-4.7$ is the same before and after scattering.}\label{f-decoherence}
\end{figure*}

The off-diagonal elements of the reduced density matrix $\rho_{cm}(r, r')$ are indicators of the degree of decoherence present in the system. To quantify the contrast between diagonal and off-diagonal elements, we introduce the function
\begin{equation}
C(r, r') = \frac{| \rho_{cm}(r,r') |}{\frac{1}{2}\left| \rho_{cm}(r, r) + \rho_{cm}(r', r') \right| }
\end{equation}
and plot it after scattering ($t = 35$) in FIG.~\ref{f-decoherence}~(a). The diagonal elements are equal to 1, and many of off-diagonal elements are only $10-20\%$ smaller than the diagonal ones, indicating large degree of coherence still present in the system.
Finally, coherence is manifested in the interference pattern when the reflected and transmitted wave packets meet at the opposite side of the ring. FIG.~\ref{f-decoherence}~(b) shows the interference pattern at the moment of maximal overlap, $t=54$; the relative coordinate degrees of freedom is traced out. The middle of the plot ($r=\pm 40$) corresponds to the point on the ring opposite to the impurity, where the reflected and transmitted wave packets meet. The bottom panel shows the same data as a density plot, simulating the interference picture observed in experiment. The interference fringe loses $\sim$15\% of its maximal visibility. For comparison, FIG.~\ref{f-decoherence}~(c) shows a similar calculation done for a single exciton scattered by the impurity. We consider an exciton wave packet with the same shape function $u_k(0)$ as in (\ref{wp-initial}) scattered by an impurity, which provides half-to-half splitting of the wave packet. Plotted is the density matrix $\rho_e(x, x)$ of exciton at the moment of the maximal overlap. The amplitude of fringe oscillations vanishes in the interference minima at panel (c), confirming that the flattening of contrast in panel (b) is due to the entanglement between relative and CM coordinates of the biexciton wavepacket.

Finally, FIG.~\ref{f-decoherence}~(d) shows the mode distribution over $K$-states in the initial wave packet (at $t = -30$), and after scattering (at $t = 35$). In \cite{bill}, decoherence due to  the excitation of internal degrees of freedom was observed. In our approximation where we only
look at the biexciton states, there is only one internal state for each $K$, namely one biexciton state. If one starts out with a wave packet with narrow distribution in $K$ then the internal states for different $K$ have a large overlap. However, due to scattering by an impurity, the overlap gets smaller as the states spread out and decoherence occurs. In the example of FIG. \ref{f-wp-dynamics} and \ref{f-decoherence}, the transmitted wave packet has more modes excited near $K\sim\pi/2$ than the reflected one has, and therefore there is some measurement of $K$ or $s$ that has some possibility of differentiating them.

\section{Discussion}
\label{s-discussion}

We have discussed two phenomena relevant to interaction of a composite particle on a lattice with a point defect. Firstly, in contrast with the case of a single exciton having only one bound state near an impurity, for biexcitons the interplay between exciton-exciton interaction and the impurity potential leads to formation of additional bound states. This can be viewed as if these two interactions introduce a finite-scale effective potential for the CM coordinate. At the same impurity strength, the number of bound states is larger when $\mbox{sgn}(V_0) = \mbox{sgn}(D)$, and the bound states form out of the scattering states near $K\sim 0$. We have also demonstrated, both analytically and numerically, the existence of bound states in the continuum for our model. Secondly, we have studied a change of the entanglement in a biexciton wave packet via one scattering event. We have observed that the entropy increases as a result of scattering. Still, we expect that the decrease of the entropy can in principle also be observed. When an initial wave packet is nearly a pure state, generally the increase of the entanglement due to a scattering event would be observed, as we saw in the previous section. However, if the initial wave packet had high entanglement of the CM motion with the internal degrees of freedom, it may drop as a result of scattering. For a free-space particle of Ref.~\cite{bill} with a quadratic dispersion, Hamiltonian can be separated into the CM and relative coordinates. In contrast, for a biexciton on a lattice possessing cosine dispersion and subject to periodic boundary conditions, the CM and relative coordinate wave-functions depend on one and the same wave number $K$, as can be seen from ~(\ref{Phi_K}). As a result, a wave packet constructed as a sum of eigenmodes is normally entangled in $r-s$-coordinate basis. We observed that while for the states with $K \sim  \pi/2$ the relative coordinate wave-functions $\phi_K(s)$ are all alike, for $K \sim 0$ and small $D$ their width strongly depends on $K$ (see FIG. \ref{f-number-of-states} for an example). If a wave packet is formed from the states near $K \sim  \pi/2$, we can approximately factorize it as $\sum_K f_K(r) \phi_K(s) \approx \phi_{K=\pi/2} (s) \sum_K f_K(r)$, which is nearly a pure state. In contrast, when $K \sim 0$ and $|D| \sim 2.1 |J|$, the states do not disentangle. This suggest a possibility to compose a highly entangled wave packet out of these low-$K$ small-$D$ states, which could purify upon propagation, by scattering with the impurity, or by other physical processes. However, as this involves small $D$, in order to study this effect one has to include two-exciton continuum states in the basis set. Such a calculation is beyond the scope of the present paper.

Our calculations are relevant for studies of quantum interference and decoherence of composite objects, as well as for the discussion of Anderson localization of interacting particles. Currently a lot of effort is devoted to understanding the role of interactions in many-body Anderson localization \cite{anderson, anderson2}, which can also be viewed as interplay between two potentials (particle--particle vs. particle--localization potential) and two length scales (particle-particle relative distance vs. the correlation length of the disorder potential). Application of our results to study of this phenomenon is underway.

Our work demonstrates important differences between lattice and free-space models. In particular, the change of the sign of the effective mass of the quasiparticle, and the resulting equivalence between the effects produced by ``attractive" and ``repulsive" potentials, have no free-space analogs. Furthermore, the property of the biexciton wave-function derived from the lattice Hamiltonian (\ref{ham}) which is inseparable in $r - s$-coordinate with periodic boundary conditions plays an important  role: The dependence of the relative coordinate wave-function $\phi_K(s)$ on the centre of mass wave vector $K$ is responsible (i) for different number of bound states for ${\rm sgn}(V_0) = \pm {\rm sgn}(D)$, and (ii) for the entanglement dynamics. Yet another difference involves resonance states. In the framework of the free-space model \cite{bill}, long-lived resonances for a composite particle at a delta-like mirror were found using complex scaling method. This method, however, becomes ineffective on a lattice, where the periodic boundary conditions rule out the very possibility of complex eigenenergies associated with Hermitian Hamiltonians \cite{scaling}. Roughly speaking, an ideal discrete periodic system is doomed to experience periodic evolution with revivals instead of exponential decay. On the other hand, it is clear that analogs of continuum resonances should exist in non-ideal lattice models, as soon as the lifetimes associated with resonances are smaller than the lifetime of the system, or when the decay length of the excitation is smaller than the system physical size. We plan to address this question in a further study.

\begin{acknowledgments}
We would like to thank I. Averbukh for attracting our attention to this problem and for discussions, and R. Krems, R. Froese, F. Queisser, O. Kabernik and T. Momose for helpful discussions and support. F. S. thanks Y. Shikano and his group at IMS for their hospitality where part of this work has been done. F. S. is partially supported by UBC international tuition award, UBC faculty of science graduate award, the research foundation for opto-science and technology, Sasakawa Scientific Research Grant.  The work was supported by NSERC Discovery grant and CFI funds for CRUCS.
\end{acknowledgments}

\appendix

      \twocolumngrid
\section{Exciton scattering by an impurity}

For an exciton scattered by an impurity in a 1D lattice, the Hamiltonian is (\ref{excitonNNA}) with free-exciton states (\ref{P(k)}) and energy (\ref{E_e}).

The Lippmann-Schwinger equation represents the total wave-function $|\psi\rangle$ as a sum of the incident state $ |\varphi \rangle $ and the scattered state, i.e., $|\psi \rangle = |\varphi\rangle + \hat{G}_0 \hat{V} |\psi\rangle$, where $\hat{G}_0 = [E_e(k) - \hat{H_0} + i\epsilon]^{-1}$ is the Green's function of the exciton and $\epsilon$ stands for any positive infinitessimal. Projecting the Lippmann-Schwinger equation onto $\bra{n}$ gives
\begin{equation}
\psi(n)=\frac{1}{\sqrt{N}}\varphi_{k}(n)+V_0~\psi(0)\sum\limits_{q}\frac{e^{iqn}}{E (k)-E(q)+i\epsilon}.
\end{equation}

\begin{figure}[b]
\includegraphics[clip,width=\columnwidth]{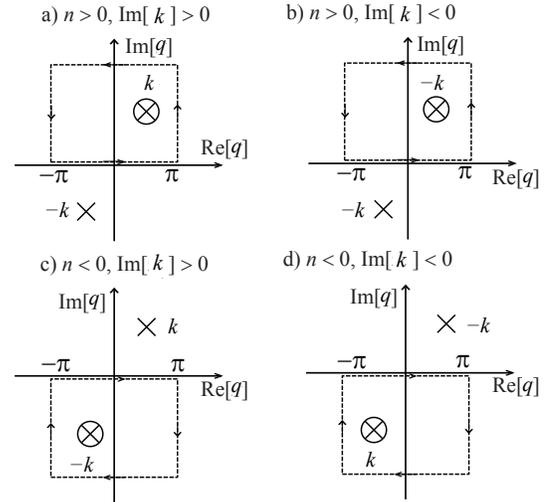}
\caption{Integration contour for various $n$ and $k$.}\label{f-contour}
\end{figure}

We find $\psi(0)$ by letting $n=0$ in this equation, and reduce it to
\begin{equation}\label{g}
\psi (n) = \frac{1}{\sqrt{N}}\left[e^{ikn}+\frac{V_0 I_{1}}{1-V_0 I_2}\right],
\end{equation}
where
\begin{eqnarray}
I_1 &=&\frac{1}{N}\displaystyle\sum_{q}\frac{e^{iqn}}{E_e(k)-E_e(q)+i\epsilon}, \nonumber\\
I_2& =&\frac{1}{N}\displaystyle\sum_{q}\frac{1}{E_e(k)-E_e(q)+i\epsilon}.\end{eqnarray}

In \cite{konobeev} this expression was examined with parabolic approximation for the dispersion of exciton, which is valid at small $k$. Here we go beyond this approximation. Replacing the sum by the integral assuming $N\to\infty$:
\begin{equation}
\displaystyle\sum_{q}\rightarrow \frac{N}{2\pi}\int^{\pi}_{-\pi}dq,
\end{equation}
and writing $E_e(k) = (E_0+2J)-4J\sin^2 (k/2)$, we get
\begin{eqnarray}
I_1=\frac{1}{8\pi J}\int^{\pi}_{-\pi}\frac{dq \ e^{iqn}}{\sin^2 (q/2)-\sin^2(k/2) - iJ\epsilon}.
\end{eqnarray}

We are interested in the complex values of $k=k'+ik''$ (${\rm Re}(k)=k'$, ${\rm Im}(k)=k''$), as they provide poles of the scattering amplitude corresponding to bound states. We choose $k' = 0$ or $\pi$ to keep the exciton energy real. Then the integral has no poles on the real axis, and $iJ\epsilon$ can be omitted. The function has two series of poles, $q = \pm k$, one laying in the upper, one in the lower half-plane. In each series the poles are shifted with respect to each other by $2\pi$. We choose the integration contour as a rectangle, with one horizontal side along the real axis segment $[-\pi, \pi]$, two vertical sides and the second horizontal line approaching $+i\infty$ if $n>0$, and $-i\infty$ if $n<0$. These contours are shown in FIG~\ref{f-contour}, along with a pair of poles $\pm k$ with $-\pi < k' < \pi$. If $k' = \pi$, the contour is shifted by $\pi$ to the right, using the periodicity of the integrand. The integrals along the right and left sides of the rectangles cancel each other, and the integral along the top side vanishes, and only one of the poles $q = \pm k$ always lays within the contour. Considering four combinations of sgn($n$) and sgn($k''$), as shown in FIG.~\ref{f-contour}, we find that the result of integration is: 
\begin{eqnarray}
I_1 =\begin{cases}
\displaystyle\frac{e^{-|k''||n|}}{2J\sinh |k''|} \qquad\mbox{ if } k'=0,\\
\displaystyle-\frac{(-1)^{n}e^{-|k''||n|}}{2J\sinh |k''|} \mbox{ if } k'=\pi.
\end{cases}
\end{eqnarray}
and $I_{2}=I_1$ with $n=0$.

Then (\ref{g}) can be written as
\begin{equation}
\psi (n) = \frac{1}{\sqrt{N}}\left[e^{ikn}+\mathcal{R}_e(k)e^{ik|n|}\right]
\end{equation}
where the scattering amplitude $\mathcal{R}_e(k)$ is
\begin{eqnarray}
\mathcal{R}_e(k)=\begin{cases}
\displaystyle \frac{V_0}{2J\sinh |k''| - V_0 }\quad \mbox{ if } k'=0,\\
\\
\displaystyle -\frac{V_0}{2J\sinh |k''| + V_0}  \mbox{ \ if } k'=\pi.\\
\end{cases}
\end{eqnarray}

The poles of this scattering amplitude coincide with those determined by (\ref{k_s}) in the limit of large $N$. Again, as in Section~\ref{s-effmass}, we note that it is the combination of signs of $J$ and $V_0$, that determines which of these two cases is realized, as $\sinh |k''|$ is always positive. When $J>0$ and $V_0 > 0$, the only pole in the reflection amplitude appears at $k=i\sinh^{-1}(V_0/2J)$. It corresponds to a single bound state marked by (i) in FIG.~\ref{f-exc}~(a) above an upper bound of the continuum spectrum with $V_0>0$. The pole at $k = \pi + i\sinh^{-1} (-V_0/2J)$ is realized when $V_0 <0$, and corresponds to a bound state (ii) in FIG.~\ref{f-exc}~(a) below the lower bound. For $J<0$, on the other hand, poles at $k=i\sinh^{-1}(V_0/2J)$ (iii) and $k=\pi + i\sinh^{-1}(-V_0/2J)$ (iv) correspond, respectively, to a bound state below the lower bound with $V_0<0$, and above the upper bound with $V_0>0$.

\section{Solving Lippmann-Schwinger equation with method of continued fractions}

In the method of continued fractions \cite{sasakawa}, the Lippmann-Schwinger equation is solved by means of iterations. In the first order,
\begin{equation}\label{Lip-Schw}
|\Psi_{K} \rangle \approx |\Phi_K \rangle + \frac{\beta_K^{(0)} \hat{G}_0 \hat{V} }{\beta_K^{(0)} - \gamma_K^{(1)}} |\Phi_K \rangle,
\end{equation}
where $|\Phi_K \rangle$ is a biexciton wave-function without the impurity (\ref{Phi_K}), $\beta_K^{(0)} = \langle \Phi_K | \hat{V} | \Phi_K \rangle$, $\gamma_K^{(1)} = \langle \Phi_K | \hat{V} \hat{G}_0 \hat{V} | \Phi_K \rangle$. 

We calculate $\gamma_K^{(1)}$ as
\begin{eqnarray}
 \gamma_K^{(1)} &=& \displaystyle\sum\limits_Q \frac{|\langle \Phi_K | \hat{V} | \Phi_Q \rangle|^2}{E_b(K) - E_b(Q) + i \epsilon} \nonumber\\
 &=& \frac{ND}{4 \pi J^2} \int\limits_{-\pi/2}^{\pi/2} \frac{dQ~ |\langle \Phi_K | \hat{V} | \Phi_Q \rangle|^2}{\sin^2 Q - \sin^2 K + i \epsilon D}.
\end{eqnarray}

For complex $K$, this integral can be carried out by the same method as in Appendix A for single exciton, with the result ($\tilde{K} = K' + iK''$):
\begin{equation}\label{B6}
\gamma_{\tilde{K}}^{(1)} = \frac{i N D |\langle \Phi_{\tilde{K}} | \hat{V} | \Phi_{\tilde K} \rangle |^2}{2 J^2 \sin 2\tilde{K}}.
\end{equation}
and the matrix element of the potential is
\begin{equation}
\langle \Phi_{\tilde{K}} | \hat{V} | \Phi_{\tilde K} \rangle  = \frac{4V_0}{N} S(K',|K''|),
\end{equation}
where
\begin{equation}\label{S}
S(K',|K''|) = \sum\limits_{s = -N/2+1}^{N/2} e^{-2|K''|s} \phi_{K'-i|K''|}(s) \phi_{K'+i|K''|}(s)
\end{equation}
is a real-valued function of $K'$ and $|K''|$. The function $S(K',|K''|)$ multiplies $V_0$, and reflects the averaging of the delta-potential by the wave function of biexciton's relative coordinate.
 Then the scattered state in (\ref{Lip-Schw}) has poles if
\begin{equation}
1 = \frac{2i  V_0 D S(K', |K''|)/J^2}{\sin(2K') \cosh(2|K''|) + i \cos(2K') \sinh(2|K''|)}.
\end{equation}

This equality with real biexciton energy can be satisfied when (a) $K' = 0$, or (b) $K' = \pi/2$. These two cases lead to two types of poles, defined, respectively, by the following equations:
\begin{equation}
\begin{array}{ll}
K' = 0, & \displaystyle  S(0,|K''|) = \frac{ J^2 \sinh(2|K''|)}{2V_0D},\\

&\\

\displaystyle K' = \frac{\pi}{2}, & \displaystyle S(\pi/2,|K''|) = - \frac{ J^2 \sinh(2|K''|)}{2V_0D}.\\
\end{array}
\end{equation}

The first type of poles correspond to the poles marked as (i),(iii) in FIG.~2~(a), the second -- to those marked as (ii) and (iv). Again, which of the two poles can be realized is determined by the signs of $V_0$ and $D$ (the latter determines the signs of the effective mass of biexciton near $K\sim 0$ and near $K\sim\pi/2$). 

Then the projection of (\ref{Lip-Schw}) onto $\langle r,s|$ can be written as
\begin{equation}
\Psi_{K} (r,s) \approx \Phi_{K}(r,s)+\mathcal{R}_b(K) \Phi_{K} (r,s),
\end{equation}
where the scattering amplitude $\mathcal{R}_b(K) $ is
\begin{equation}
\mathcal{R}_b(K) = \frac{2i D V_0 S(K',|K''|)}{J^2 \sin2(K'+i|K''|)-2iDV_0S(K',|K''|)}.
\end{equation}

\end{document}